# High-responsivity vertical-illumination Si/Ge uni-traveling-carrier photodiodes based on silicon-on-insulator substrate

Chong Li, ChunLai Xue, Zhi Liu, Hui Cong, , Buwen Cheng, Xia Guo, and Wuming Liu

**Si/Ge uni-traveling carrier photodiodes exhibit higher output current when space-charge effects are overcome and thermal effects are suppressed, which is highly beneficial for increasing the dynamic range of various microwave photonic systems and simplifying high-bit-rate digital receivers in different applications. From the point of view of packaging, detectors with vertical-illumination configuration can be easily handled by pick-and-place tools and are a popular choice for making photo-receiver modules. However, vertical-illumination Si/Ge uni-traveling carrier (UTC) devices suffer from inter-constraint between high speed and high responsivity. Here, we report a high responsivity vertical-illumination Si/Ge UTC photodiode based on a silicon-on-insulator substrate. The maximum absorption efficiency of the devices was 2.4 times greater than the silicon substrate owing to constructive interference. The Si/Ge UTC photodiode was successfully fabricated and had a dominant responsivity at 1550 nm of 0.18 A/W, a 50% improvement even with a 25% thinner Ge absorption layer.**

**Introduction**

1High-current photodiodes, which receive communication signals in the near-infrared range, are highly beneficial in various photonic systems for increasing their dynamic range[1][2] and simplifying high-bit-rate digital receivers[3]. The output radio-frequency signal level from such photodiodes can be increased with the response photocurrent, and are thus a particularly important component for optically-steered phased array antennas, which can help the antenna to reduce its phase- and amplitude-matched electronic gain [4,5,6]. However, the conventional pin structure has a limitation in current density during high frequency operation, owing to the space-charge effect [7,8].

The uni-traveling carrier (UTC) structure was designed to overcome the space-charge effect and increase the transition frequency using a p-type doped absorption layer instead of a conventional intrinsic layer [9,10,11,12,13,14]. However, the output power of these devices was further limited by thermal effects [15,16]. Monoatomic crystals of materials such as Ge and Si material have higher thermal conductivity than that of InGaAs and InP alloy materials [17]. Additionally, Si/Ge devices have great advantages in their compatibility with complementary metal-oxide-semiconductor (CMOS) technology and large-scale monolithic integration circuits, low cost, and low power consumption [18,19,20,21,22]. Therefore, Si/Ge uni-traveling carrier photodiodes have dramatic practical potential for high-current output applications [23,24]. Besides, from the point of view of packaging, detectors having a vertical-illumination configuration can be easily handled by pick-and-place tools and are consequently a popular choice for making photo-receiver modules [25,26]. Therefore, the most commonly-used photodetectors are of the vertical-illumination type. To our knowledge, the best performance obtained for a vertical-illumination Si/Ge UTC device until now was reported by M. Piels, who demonstrated a low thermal impedance of 520 K/W and a 1 dB saturation photocurrent of 20 mA. However, the responsivity of this device at 1550 nm was as low as 0.12 A/W with a 0.8-μm-thick Ge absorption layer [24]. Such a low responsivity could seriously increase power dissipation and limit high-output applications. Although increasing the thickness of the Ge absorption layer could improve the responsivity of the device according to $R \propto (1-e^{-aD})$ [27], the electron transit time also increases with Ge thickness, which decreases device response speed [28]. Therefore, it is a challenge to obtain both high responsivity and high speed at the same time in vertical-illumination Si/Ge UTC detectors.

Silicon-on-insulator (SOI) substrates have great advantages that can improve the responsivity and bandwidth performance of Si/Ge photodiodes. First, the large difference in refractive index between the buried oxide layer (BOX) and the Si is beneficial in recycling transmission light back to the absorption layer, which is equivalent to extending the absorption length. This allows the absorption efficiency of the photodiodes to be increased without sacrificing the response speed. Second, high quality Ge film with low threading dislocation density (TDD) can be obtained on the top silicon membrane of SOI by elastic deformation, and adapt to the lattice of a hetero-epitaxial file grown upon it [29,30]. The threading dislocations inside the Ge layer can decrease carrier lifetime and increase the non-radiative recombination rate [31,32], which can reduce the number of photon-generated carriers collected by the metal contacts. Therefore, the carrier collection efficiency of the device may be enhanced using a SOI substrate. Third, use of a SOI substrate could reduce the parasitic capacitance of the device, which would result in an improvement in frequency response performance [33,34] and a decrease in power loss [35].

Here, we report a high-speed, high responsivity vertical-illumination Si/Ge UTC-PD based on a silicon-on-insulator (SOI) substrate. The silicon-on-insulator substrate was used to reflect transmission light for high absorption efficiency, and to improve the lattice quality of the Ge epitaxial layer to increase the efficiency of photon-generated carrier collection. The absorption efficiency of the Ge-on-SOI UTC photodiode was found to vary periodically with the thicknesses of both the BOX and Si layers, owing to the interference between the incident light and the light reflected by the BOX layer of the SOI. Moreover, the maximum absorption efficiency of the devices on SOI was found to be 2.4 times greater than that of the silicon substrate and 4.9 times greater than the minimum absorption efficiency. Si/Ge UTC photodiodes on SOI substrate with a 0.6-μm-thick Ge absorption layer were fabricated and characterized. The responsivity of the photodiodes at 1550 nm was improved to 0.18 A/W.

The −1-dB compression current of the 15-μm-diameter device was 16.2 mA at 3 GHz, with a 3-dB bandwidth of 9.73 GHz.

**Results**:

**Structure and electric field**. Figure 1(a) shows a cross-sectional schematic view of a Si/Ge UTC photodiode based on a commercially available SOI substrate with 1.0-μm-thick n-doped Si and 2-μm-thick BOX layers. A step gradient doping profile was employed, which enabled the generation of several regions with high local electric field to further decrease the transit time of the photo-generated electrons, as the red curve shown in Fig. 1(b). A simulated band-gap diagram and electric field distribution at 0 V of such a device are illustrated in Fig. 1(b) by the blue and red curves, respectively, calculated after modifying the doping parameters according to the results of secondary ion mass spectrometry (SIMS) measurements. The built-in electric field was generated from the differences in the doping concentration. Each abrupt change in doping concentration corresponded to an electric field peak. The photon-generated electrons were accelerated in the Ge absorption layer and gained kinetic energy to pass through the Si/Ge heterojunction barrier under the action of the built-in electric field. A larger electric field and thus lower transit time can be obtained compared with those achievable with conventional linear gradient doping of the absorption layer [36,37,38]. A micrograph of a photodiode with a 15-μm-diameter top mesa is shown in the inset of Fig. 1(a).

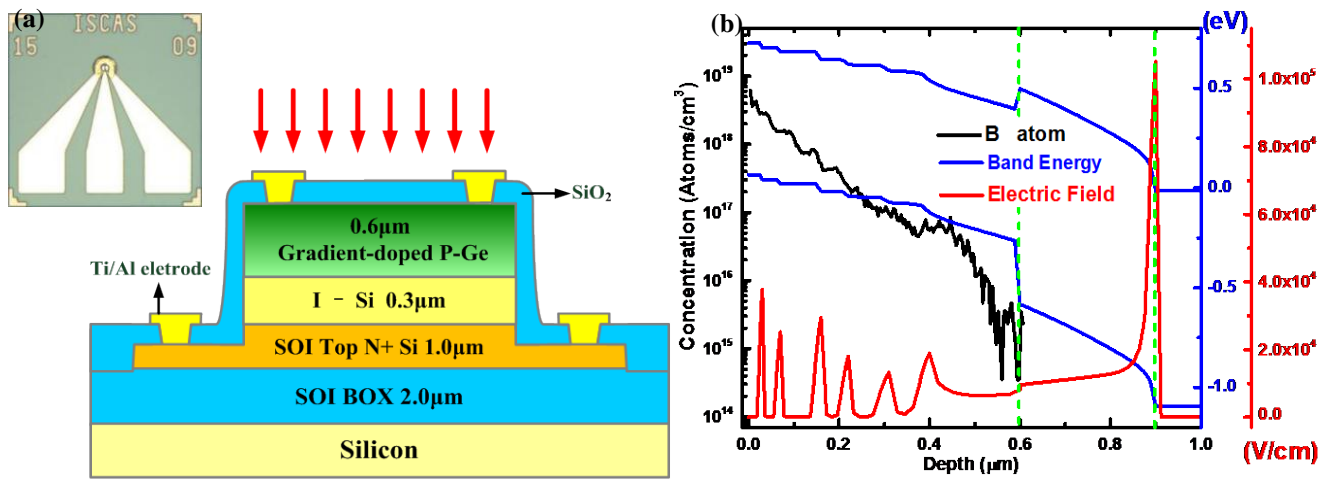

Fig. 1 (a) Cross-sectional schematic view of the reported Ge-on-SOI UTC photodetector and top view of a double-mesa structure of the Ge-on-SOI UTC photodetector. The substrate was SOI with a 1.0-μm-thick n-doped top Si film and a 2-μm-thick BOX layer, the collect layer was a 0.3-μm-thick intrinsic epitaxial silicon layer, and the absorption layer was a 0.6-μm-thick epitaxial germanium layer with step gradient doping of B atoms. (b). The left black coordinate and curve show that the doping concentration of B atoms in the Ge absorption layer step decreased from $5 \times 10^{19}$ to $2 \times 10^{17}/cm^3$, as determined by SIMS. The etch step was nearly 0.05 μm wide, resulting in six high local electric fields to accelerate the photon-generated electrons and shorten the transmit time. The right red and blue coordinates show the electric field and band energy of our devices without bias, respectively. The peak value and width of the six local electric fields were determined by the doping concentration around the step interfaces. According to our simulation, six step gradients in the absorption layer enabled a maximum potential difference across the layer.

**Responsivity characterization.** The responsivity of a vertical-illumination photodiode is limited mainly by a combination of three factors: (1) the coupling efficiency determined by the top anti-reflection coating; (2) the absorption efficiency of the Ge layer; (3) the collection efficiency of the photon-generated carriers [39]. The former two factors can be optimized through the structural design of the devices. The last one is mainly determined by the quality of the epitaxial crystalline Ge and the electric field inside the devices. Generally, only the light coupled into the absorber ($P_0$) can be converted into electron-hole pairs. To maximize the coupled light, the thickness of the top anti-reflection coating should be N·(λ/4n), according to destructive coherence inside the coating, where N is a positive integer, and n is the refractive index of the coating [40]. The carrier collection efficiency of Si/Ge UTC photodiodes is mainly determined by the design of the electric field inside the devices and by the recombination caused by defects inside the Ge layer and at the hetero-interface [41,42].

Absorption efficiency is generally dependent on the absorption coefficient and thickness of the absorption layer. The absorption coefficient of Ge is relatively low at 1550 nm, which is near the band-gap edge. The power inside the absorption layer $P_{ab}$ can be expressed by $P_{ab}=P_0(1-e^{-aD})$, where $α$ is the absorption coefficient of the absorber and $D$ is thickness of the absorber. The introduction of the SOI substrate was expected to cause recycling of the transmission light and improve the light absorption of the device. The new absorption power is:

$$P_{ab}=P_0(1-e^{-aD}) + P_0 \cdot e^{-aD} \cdot R_{ref} \cdot (1-e^{-aD}) \tag{1}$$

where $R_{ref}$ is the reflection coefficient of the SOI substrate, and $R_{ref}$ is the reflection coefficient of the top coating film and Ge film. The detailed optical power distributions in the Si/Ge UTC photodiodes on the Si and SOI substrates were compared by simulation with the commercial finite-difference-time-domain (FDTD) simulation package, as shown in Fig. 2. The scale bar illustrates the optical power. The optical power inside the Si bottom layer of the SOI substrate is obviously much lower than that in the silicon substrate, which indicates the unemployed or transit light power of the devices. Therefore, SOI substrate is more beneficial to higher light absorption by Ge through reflection of the BOX compared with the Si substrate.

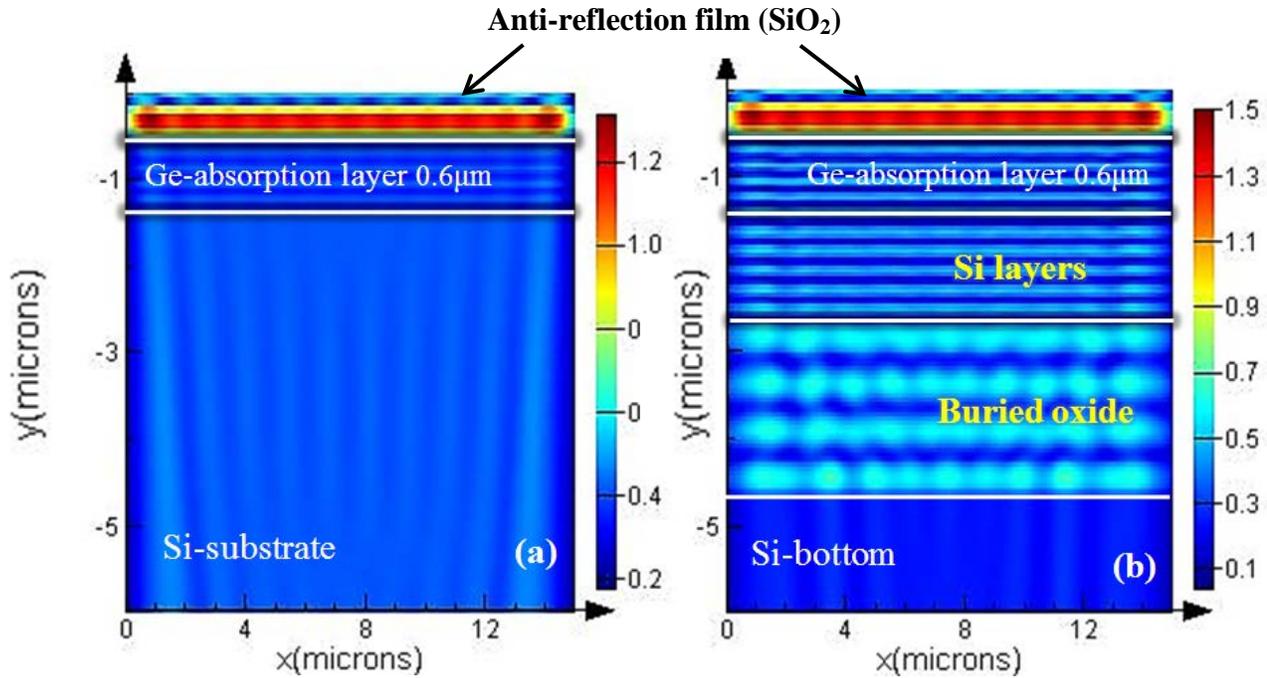

Fig. 2. Optical power distribution inside Si/Ge UTC devices grown on (a) Si substrate and (b) SOI substrate. The only difference between the two devices was the inserted 2-μm-thick BOX layer, and the thickness of the silicon collection and contact layers was 1.3 μm. The scale bar illustrates the optical power. Because of the coherence effect between the incident light and reflected light of the BOX layer, the optical power inside the Ge, Si, and BOX exhibited a periodic enhancement distribution. The period was determined by the refractive index and the wavelength of the incident light. The light inside the Si bottom layer of the SOI substrate and inside silicon substrate was the unemployed light of the devices. Obviously, that in the SOI was much lower than that in the silicon substrate. Thus, the SOI substrate was more beneficial for higher light absorption by Ge, compared with the Si substrate.

The relationship between the absorption efficiency and the thickness of the BOX and silicon layer was analysed and calculated according to the SMM using Matrix Laboratory (MATLAB). When the thickness of the anti-reflection coating was fixed, the absorption efficiency varied periodically with the thicknesses of both BOX and Si, shown in Fig. 2. The absorption efficiency of the device without BOX was 0.092 with a 0.6-μm-thick coating film, but the maximum obtained was 0.217 with a 1.1-μm-thick Si layer and a 1.85-μm-thick BOX, more than double. However, the general thickness of the BOX layer in commercial SOI substrates is 2 μm. Because of the constructive interference between the reflected and input waves, the thickness period ($T$) is $T = \lambda/2n$, where $\lambda$ is the input wavelength and $n$ is the refractive index of the transmission media. Here, the silicon thickness cycle was 0.223 μm with a Si thickness of 1.3 μm, i.e., the structural parameters of our device as shown in Fig. 1(a). Thus, theoretically, the maximal absorption efficiency is about 0.168, and the ideal responsivity is 0.208 A/W without photon-generated carrier recombination.

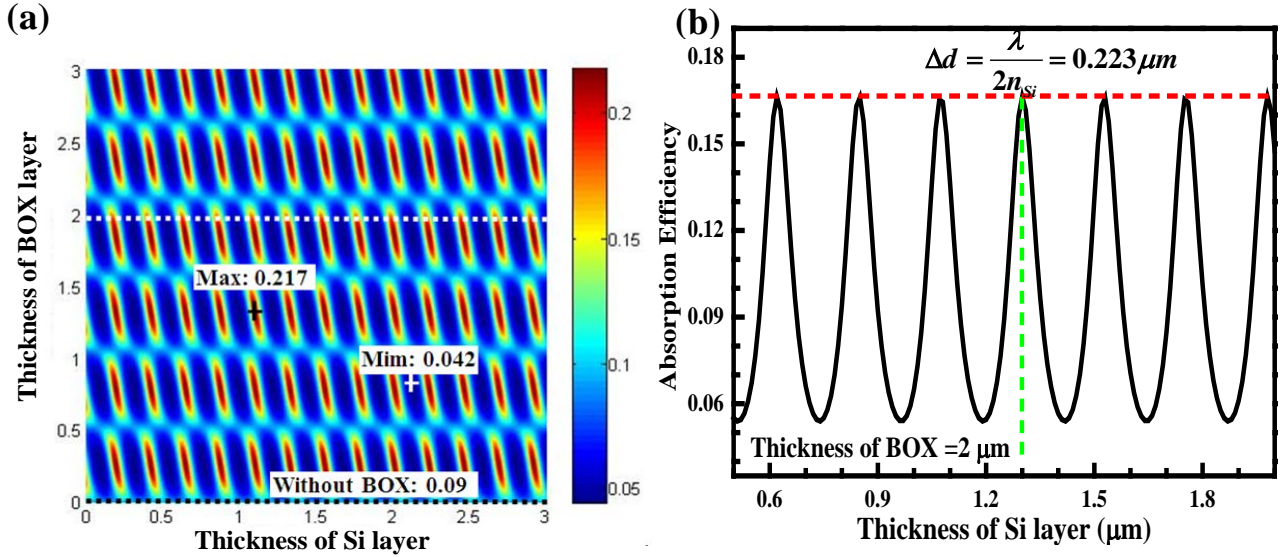

Fig. 3 The wavelength of the incident light was 1550 nm, and the absorption efficiency of the Ge material was 1000/cm. (a) Relationship between the absorption efficiency of Ge-on-SOI UTC photodiode and the thicknesses of the BOX and silicon layers. Using these thicknesses of the BOX and silicon as ordinate and abscissa, respectively, the absorption efficiency is mapped in colored points in a two-dimensional coordinate plane. The bottom black dashed line represents the absorption efficiency of the devices without BOX layer, which was 0.09. The other two special points are the maximum of 0.217 with a 1.1-μm-thick Si layer and a 1.85-μm-thick BOX, and the minimum of 0.044 with a 2.13-μm-thick Si layer and a 1.32-μm-thick BOX. (b) Relationship between the absorption efficiency and the thickness of the silicon layer, when the thickness of the BOX is the general commercial value of 2 μm. The absorption efficiency changes periodically with the thickness of Si layer because of constructive interference. The period is $T = \lambda/2n$, which is 0.223 μm for the silicon layer, and the maximum efficiency is 0.168 when the silicon thickness is 1.3 μm, which is the structural parameter of our device.

The experimentally determined device current for the 15-μm-diameter photodiodes, without illumination and with the normally incident light on the top surface, is shown in Fig. 4. The dark current was 58 nA under a reverse bias of 1 V, which corresponds to a current density of 96.3 mA/cm$^2$. The minimum dark current density was approximately 61.9 mA/cm$^2$ for the 40-μm-diameter device at −1 V. The dark current could be further reduced by appropriate thermal processing to decrease the threading dislocation density around the Si/Ge interface[11][12], by the passivation process, or by the application of a guard-ring around the sidewall.

At a reverse bias of 1 V, the optical responsivity was 0.18 A/W at 1550 nm with a 0.6-μm-thick Ge layer, 50% higher than that of the existing Si/Ge UTC devices, which is $R$=0.1 A/W with a 0.8-μm-thick Ge layer[7]. In Fig. 4, the mismatch between the dark and optical currents between 0.7 V and 0.3 V arose from the diffusion and collection of the photo-generated carriers because of the gradient doping of B and P atoms in the actual epitaxial layers. Because of the carrier recombination, this measurement responsivity is little lower than the former theoretical results.

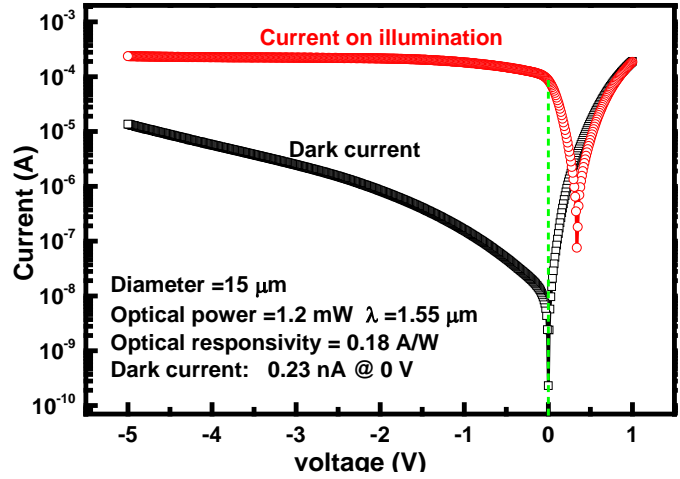

Fig. 4. Current-voltage characteristics of the 15-μm-diameter device without illumination and under laser irradiation with an input optical power of 1.2 mW at 1550 nm. The dark current without bias was 0.23 nA. When the reverse bias was increased to 1 V, the dark current rose to 58 nA, corresponding with a current density of 96.3 mA/cm. The optical responsivity was 0.18 A/W at 1550 nm. The dashed line indicates that there was a saturation of the optical responsivity values at 0 V bias, which indicates that this photodetector configuration allowed nearly complete photo-generated carrier collection without bias.

**3-dB bandwidth characterization.** The bandwidth of common photodiode is limited mainly by the resistor–capacitor (RC) bandwidth ($f_{RC}$) and the carrier transit-time-limited bandwidth ($f_t$) in the active region [43]. Particularly, for UTC devices, the most important limitation is the transfer time of electrons in the p-type absorption layer owing to their low diffusion velocity in it. Based on the principle of conservation of energy, the electrical potential difference across the absorption layer determines the change in kinetic energy of the carriers. A high kinetic energy in the absorber could shorten the transfer time of the electrons. Based on the direct relationship between the doping difference across the doped junction and the electrical potential difference, a step gradient-doping region was introduced in the absorption layer to increase the potential difference across the Ge layer. We can assume that the photo-generated electrons drifted with saturation velocity ($v_s$) in the p-type absorption layer of our devices owing to the high potential drop across the absorber, and then across the collection layer with thermionic emission velocity ($v_{th}$), because of the effect of the thermionic emission [6]. Then, the carrier transit frequency can be approximated by the following equation:

$$f_t = 1/(2\pi\tau_a) = 1/(W_a/v_s + W_c/v_{th}) = 1/(W_a/v_s + W_c/\sqrt{2kT/\pi m_e^*}), \qquad (2)$$

where $v_{th}$ is the thermionic emission velocity, $v_s$ is the saturation velocity, and $m_e^*$ is the effective mass of the electrons. The capacitor in the absorption layer is ignored, so $f_{RC}$ can be approximated using:

$$f_{RC} = \frac{1}{2\pi(R_s + R_L)C} = \frac{1}{2\pi(R_s + R_L)\cdot\dfrac{\varepsilon\varepsilon_0 \pi D^2}{4W_c}}, \qquad (1)$$

where $W_c$ is the collection layer thickness, $W_a$ is the absorption layer thickness, $D$ is the mesa diameter, $R_L$ is the load resistance (50 Ω in this case), $R_S$ is the series resistance, and $\varepsilon$ and $\varepsilon_0$ are the relative and vacuum permittivity, respectively. If the series resistance of the device is about 100 Ω, which was the measured resistance of our devices equal to the slope of I-V curve at the positive bias of nearly 0.26 V [44]. The theoretical values of $f_{3dB}$ with various diameters are shown in Fig. 5(a). The results are almost consistent with the experimental results except for the 15-µm-diameter device, as shown in Fig. 5(b). Therefore, the step gradient-doping design was able to make the carriers drift in the absorber with a saturation velocity and efficiently increased the transit frequency of the UTC photodiodes. The different result observed for the 15-µm-diameter device may be from the higher series resistance of smaller size devices caused by the fabrication processes.

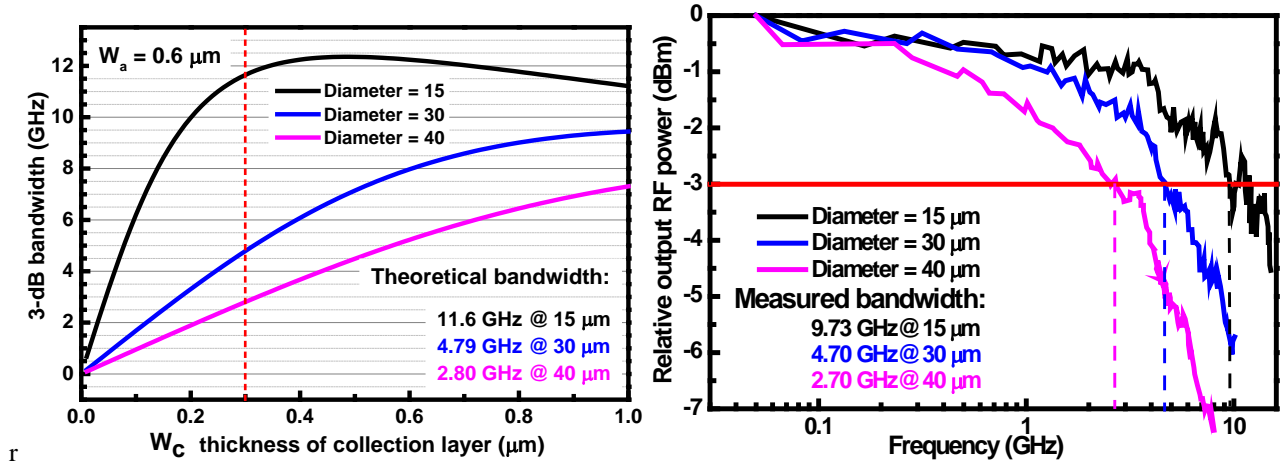

Fig. 5. Frequency responses for UTC photodiodes with diameters of 15, 30, and 40 µm under 1550 nm incident light: (a) theoretically predicted values for a series resistance of 100 Ω, which was measured using the slope of the I-V curve at the positive bias of ca. 0.26 V and on the assumption that the photo-generated electrons traveled across the p-type absorption layer by drifting with saturation velocity ($v_s$), and across the collect layer by drifting with thermionic emission velocity ($v_{th}$). (b) The 3-dB bandwidth was measured with a vector network analyser. The theoretical $f_{3dB}$ values are almost consistent with the experimental results except for the 15-µm-diameter device, whose difference may have resulted from the higher series resistance of smaller size devices caused by the complicated fabrication processes.

**Saturation characterization.** The device saturation current was obtained using large signal measurements, as shown in Fig. 6. A 100% modulation depth tone was fixed at 3 GHz for measurement of the 15-µm-diameter device. The fabricated devices exhibited high saturation photocurrents. The 1-dB compression currents were measured to be 16.2 mA at reverse bias voltages of −6 V. For the 40-µm-diameter device with a 3 dB bandwidth of 2.7 GHz, the saturation current was 16.24 mA with a fixed modulation frequency of 1 GHz under −7 V bias, and the RF output power was 4.6 dBmW. The saturation of the Ge-on-SOI UTC photodiodes could be further improved by suppressing thermal effects [45,46] by substrate thinning and decreasing the thickness of the BOX layer.

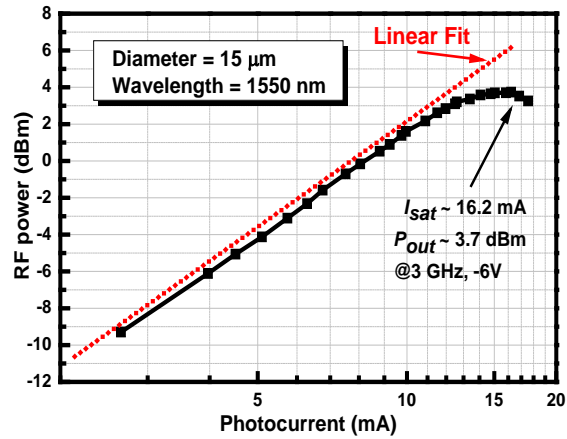

Fig. 6. Results of large signal −1-dB compression photocurrent measurement for the 15-µm-diameter Si/Ge UTC photodiodes. The incident light had a wavelength of 1550 nm, 100% modulation depth, and modulation frequency fixed at 3 GHz. The saturation current was 16.2 mA at a reverse bias of 6 V, and the output RF power was 3.7 dBmW. The saturation current and RF power were further increased with the reverse bias.

**Discussion**

The demonstrated on-chip performance of the present high-responsivity vertical-illumination Si/Ge uni-traveling-carrier photodiodes paves the way for all kinds of vertical-illumination Si/Ge photodetectors with high responsivity, and high-quality epitaxial germanium. It will also allow the realization of large-scale monolithic integrated microwave optoelectronic antenna systems with low cost and low power consumption. The use of silicon-on-insulator substrate for Si/Ge UTC photodiodes offers advantages in reflecting the transmission light to increase the absorption efficiency of the input optical signal, and improving the lattice quality of Ge epitaxial layer to increase the efficiency of photon generated carrier collection. Because of the constructive interference between the incident light and the light reflected by the buried oxide layer of the SOI, the maximum absorption efficiency of the devices on SOI substrate was 2.4 times greater than that obtained with silicon substrate and 4.9 times greater than the minimum absorption efficiency. The photodiodes showed a responsivity of 0.18 A/W at a wavelength of 1.55 µm, which is a 50% higher responsivity with a 25% thinner Ge absorber than that reported for previous Si/Ge UTC devices. Furthermore, use of a step gradient-doping absorber caused the carriers to drift with a saturation velocity in the absorber, which efficiently increased the transit frequency of the UTC photodiodes. As a result, the 3-dB bandwidth of the 15-µm-diameter device was improved to 9.72 GHz under a −5 V bias voltage. The 1-dB compression current of the device was 16.2 mA at 3 GHz.

**Methods**

**Silicon and germanium film growth and characteristics.** After the growth of the intrinsic Si layer at 750 °C by cold-wall ultra-high vacuum chemical vapour deposition on the SOI substrate using a source gas of pure $Si_2H_6$ (UHV-CVD), a brief interruption was introduced to decrease the growth temperature to 290 °C for the growth of the 60-nm-thick p-doped Ge buffer layer. A 600-nm-thick boron-doped Ge layer was then grown on the top as the absorption layer at 600 °C using pure $GeH_4$ and diluted $B_2H_6$ source gases. Six boron-doping concentration steps were made in the Ge absorption layer, decreasing from $5 \times 10^{19}$ to $2 \times 10^{17}/cm^3$, which was confirmed by SIMS measurements, as shown by the black curve in Fig. 3(b).

**Fabrication and characterization of photodiodes.** Circular Ge layer mesas for normal incidence Si/Ge UTC photodiodes with diameters ranging from 15 to 40 μm were defined by standard photolithography and inductively coupled plasma (ICP) etching. The second mesa was etched to the 2-μm-thick buried oxide layer. The double mesa layout significantly reduced the parasitic capacitance. Top and bottom contacts were lithographically defined on evaporated Ti/Al and a rapid-thermal-annealing (RTA) process was carried out for impurity activation. A passivation/antireflection coating was deposited by plasma enhanced chemical vapour deposition (PE-CVD). Windows for the metal contacts were opened by $C_4F_8$ ICP etching. The metal pad was evaporated and lifted off. A micrograph of a photodiode with a 15-diameter top mesa is shown in Fig. 1(b). The current-voltage characteristics of our device was measured using an Agilent B1500A semiconductor parameter analyser on a probe station at room temperature. The photocurrent-voltage characteristics were obtained under laser irradiation at a wavelength of 1550 nm with power of 1.2 mW.

**Saturation measurements.** The device saturation current was obtained using large signal measurements. A heterodyne technique using two free-running lasers at 1550 nm was used and a modulation index was ultimately obtained. A 100% modulation depth tone was fixed at −3 GHz for measurement of the 15-μm-diameter device.

**Acknowledgements**

This work was supported in part by the Major State Basic Research Development Program of China under grant nos. 2013CB632103 and 2011CBA00608, by the National High-Technology Research and Development Program of China under grant no. 2012AA012202, and by the National Natural Science Foundation of China under grant nos. 61222501, 61335004, 11434015, and 61227902, and the Specialized Research Fund for the Doctoral Program of Higher Education of China under grant no. 20111103110019.